\documentclass[aps,pra,twocolumn,showpacs,superscriptaddress,]{revtex4-1}
\usepackage{amssymb,amsmath}
\usepackage{graphicx}
\usepackage{dcolumn}
\usepackage{bm}
\usepackage{color}

\begin{document}

\title{Luttinger parameter of quasi-one-dimensional para-H$_{\bf 2}$
}

\author{G. Ferr\'e}
\affiliation{Departament de F\'{i}sica, Universitat Polit\`{e}cnica de 
Catalunya, Campus Nord B4-B5, E-08034 Barcelona, Spain}

\author{M. C. Gordillo}
\affiliation{Departamento de Sistemas F\'{\i}sicos, Qu\'{\i}micos y 
Naturales, Universidad Pablo de Olavide, E-41013 Sevilla, Spain}

\author{J. Boronat}
\affiliation{Departament de F\'{i}sica, Universitat Polit\`{e}cnica de 
Catalunya, Campus Nord B4-B5, E-08034 Barcelona, Spain}

\email{jordi.boronat@upc.edu}

\begin{abstract}
We have studied the ground-state properties of para-hydrogen in one dimension 
and in quasi-one-dimensional configurations using the path integral ground 
state Monte Carlo method. This method produces zero-temperature exact results 
for a given interaction and geometry. The quasi-one-dimensional setup has been 
implemented in two forms: the inner channel inside a carbon nanotube coated
with H$_2$ and a harmonic confinement of variable strength. Our main result is 
the dependence of the Luttinger parameter on the density within the stable 
regime. Going from one dimension to quasi-one dimension, keeping the 
linear density constant, produces a systematic increase of the Luttinger 
parameter. This increase is however not enough to reach the superfluid regime 
and the system always remain in the quasi-crystal regime, according to 
Luttinger liquid theory.
\end{abstract}


\maketitle

\section{Introduction}
\label{sec:introduction}

The search for a superfluid phase in molecular \textit{para}-hydrogen 
(\textit{p}-H$_2$) started from the theoretical proposal by Ginzburg and 
Sobyanin in 1972~\cite{Ginzburg72}. They suggested that  \textit{p}-H$_2$, with 
spin 1, should be superfluid under a transition temperature $T_\lambda$ that 
they estimated to be $T_\lambda \sim 6$ K using ideal Bose gas theory. This 
relatively high temperature, compared with the well-known transition temperature 
in $^4$He ($T_\lambda(^4\text{He})=2.17$ K), was the result of the smaller mass 
of \textit{p}-H$_2$. However, this estimation is too crude because the strong 
interactions between the \textit{p}-H$_2$ molecules are simply ignored. 
Moreover, the transition temperature for the ideal Bose gas increases with the 
density $\rho$ as $\rho^{2/3}$ whereas it is known that in superfluid $^4$He it 
slightly decreases with $\rho$. Later on, Apenko~\cite{Apenko99} proposed a 
phenomenological theory similar to the Lindemann criterion for classical crystal 
melting. He concluded that in \textit{p}-H$_2$ $T_\lambda$ should vary between 
1.1 and 1.2 K, depending on the density. A recent path integral Monte Carlo 
(PIMC) simulation of    \textit{p}-H$_2$ at low temperatures, in which it was 
possible to frustrate the formation of the stable crystal, showed that 
superfluidity appears at temperatures around 1 K~\cite{osychenko}.      

Superfluidity in bulk hydrogen is not observed because it crystallizes in an 
hcp phase at a temperature $T=13.8$ K which is much higher than the estimated 
transition temperature $T_\lambda$. The mean reason is that the intermolecular 
interaction is around three times more attractive than the one between He 
atoms. 
This enhanced attraction dominates over the positive effect produced by  
the smaller mass of H$_2$ respect to the $^4$He one. There have been a number 
of supercooling attempts to create a metastable liquid phase but even at $T\sim 
9$ K the liquid phase freezes quickly into a crystal~\cite{kuhnel11}. One of the 
a priori more interesting options was to confine hydrogen in a porous media, 
like a vykor glass, with pores in the nanometer scale. However, the lowest 
temperature at which the system was detected to be liquid was $T\sim 8$ K and so 
still far from the pursued superfluid~\cite{sokol}.

At present, the only experimental signatures of superfluid \textit{p}-H$_2$ 
come from experiments with small doped clusters~\cite{Grebenev00}. By measuring 
the rotational spectra of the embedded molecule it was possible to determine the 
effective moment of inertia of the cluster and thus the superfluid fraction. 
These experiments show significant evidence of superfluidity in clusters made up 
of $N \le 18$ molecules. Larger clusters of up to $N\sim 10^4$ molecules down to 
a temperature $T=2$ K have recently been produced but with no signature of 
superfluidity due to this still too high temperature~\cite{Prozument08}. 
Another way of 
frustrating the formation  of the crystal was the generation of continuous 
hydrogen filaments of macroscopic dimensions, but again without signature of 
superfluidity~\cite{Grisenti06}. 

On the theoretical side, the search for superfluidity in \textit{p}-H$_2$ has 
been intense in the last decades. The well-known radial interaction between the 
molecules and the progress achieved in quantum Monte Carlo methods have allowed 
for accurate results in different geometries. To frustrate the crystal 
formation and reduce the strength of the interactions it was proposed to work 
with a two-dimensional geometry with some impurities arranged in a periodic 
lattice~\cite{Gordillo97,claudi1}. First results obtained within this scheme 
found finite superfluid densities but posterior simulations were not able to 
reproduce these signatures~\cite{Boninsegni05H2,claudi2,Bonin_na}. The greatest 
effort was devoted to the study of small clusters, both 
pure~\cite{Sindzingre91,Mezzacapo06,Mezzacapo07,
Khairallah07,Mezzacapo08,ester,guardiola,cuervo06,cuervo08,cuervo09} and doped 
with impurities~\cite{Kwon02,
Paesani05,Kwon05}. There is an overall consensus that  \textit{p}-H$_2$ becomes 
superfluid at temperatures smaller than 1-2 K and that the superfluid fraction 
decreases fast with the number of molecules of the cluster. For $N > 18$-25 the 
superfluidity vanishes and solid-like structures are observed. 

Recently, there has been interest in the study of  \textit{p}-H$_2$ in 
quasi-one-dimensional environments~\cite{gordillo2,ancilotto,bonintubo}. Again, 
the idea is to reduce dimensionality to soften the intermolecular attraction. 
Quantum Monte Carlo calculations of hydrogen adsorbed inside narrow pores of 
different size and nature have been performed showing, in some cases, the 
existence of inner channels which behave as effectively one-dimensional 
systems. Interestingly, a recent ground-state quantum Monte Carlo 
calculation~\cite{ancilotto} 
has shown that the inner channel of  \textit{p}-H$_2$ adsorbed inside a (10,10) 
armchair carbon nanotube is superfluid.

In the present work, we study the one-dimensional character of narrow channels 
of  \textit{p}-H$_2$ and determine the Luttinger 
parameter~\cite{Haldane,Cazalilla,Giamarchi} as a function of the linear 
density. Our method is the path integral ground state (PIGS), a 
zero-temperature 
approach  which is able to generate exact results for the ground-state of the 
system~\cite{sarsa}. We have studied three different cases: a purely 1D array 
of molecules,  \textit{p}-H$_2$ inside a (10,10) carbon nanotube coated with 
an incommensurate layer of hydrogen, and  \textit{p}-H$_2$ confined harmonically 
to move in a channel of different widths. Our results show that moving from 1D 
to quasi-1D reduces effectively the interaction producing an increase of the 
Luttinger parameter. However, this slight increment is not enough to achieve 
the superfluid-like behavior within Luttinger theory. The system breaks its 
homogeneity when crossing the spinodal point and this happens clearly before of 
getting superfluity, in contradiction with the recent findings of 
Ref.~\cite{ancilotto}. 

The rest of the paper is organized as follows. In the next Section we briefly 
discuss the method used in our analysis and describe the properties of the 
three different studied setups. Our results are reported in Sec. III, mainly 
the dependence of the Luttinger parameter with the density, within the 
stability 
regime of the quasi-1D system. Finally, an account of the main conclusions is 
given in Sec. IV.

\section{Method}
\label{sec:method}

The ground-state energy and structure properties of quasi-one-dimensional 
\textit{p}-H$_2$ have been studied using the path integral ground state (PIGS) 
method~\cite{sarsa}. For bosons and a given interaction, this method is exact 
within controlled statistical noise. The ground-state wave function of the 
$N$-body 
system is obtained from
\begin{equation}
\Psi(\bm{R}) = \int \, G(\bm{R},\bm{S},\tau) \psi_{\text{m}}(\bm{S}) d \bm{S} \ 
,
\label{pigswf}
\end{equation}
with $\bm{R}=\{\bm{r}_1,\ldots,\bm{r}_N\}$,   $G(\bm{R},\bm{S},\tau)$ the 
Green's function in imaginary time $\tau$, and 
 $\psi_{\text{m}}(\bm R)$ a model wave function with the proper Bose symmetry.
Obviously, the Green's function for a generic time $\tau$ is not known in 
general but one can build $G$ from its knowledge at short times and then apply 
its convolution property to arrive to the desired total time $\tau$. This is, 
in fact, the same method used at finite temperature (the path integral 
Monte Carlo method (PIMC)) just changing the 
imaginary time to inverse of temperature and closing the paths instead of 
being open as in PIGS. 

In our simulations, we approximate the Green's function at short time using the 
fourth-order splitting proposed by Chin and Chen~\cite{chin}, that in previous 
calculations has shown high accuracy~\cite{sakkos}. The trial wave function  
$\psi_{\text{m}}(\bm{R})$ in Eq. \ref{pigswf} plays the role of boundary 
condition of the open paths. As we simulate a Bose system it has to be 
symmetric under exchange of particles but its specific shape is rather 
irrelevant, its effect being mainly on the total imaginary time to project 
out the ground-state wave function~\cite{rota}. In the present work, we have 
used a Jastrow model with McMillan correlation factors, 
$\psi_{\text{m}}(\bm{R})=\prod_{i<j} \exp(-0.5 (b/r_{ij})^5)$, with $b=3.71\ 
\text{\AA}$. It is worth noticing that PIGS provides pure (unbiased) 
estimators of diagonal and non-diagonal operators $\hat{O}$ by calculating them 
in 
the center of a symmetric chain with $\psi_{\text{m}}$ at both extremes, that 
is,
\begin{equation}
\langle \hat{O} \rangle = {\cal N}^{-1} \, \langle \psi_{\text{m}} | G(\tau) \, 
\hat{O} \, G(\tau) | \psi_{\text{m}} \rangle \ ,
\label{pure}
\end{equation}
with $\cal N$ a normalization constant.

The Hamiltonian of the system is
\begin{equation}
H= - \frac{\hbar^2}{2m} \sum_{i=1}^{N} {\bm \nabla}_i^2 + \sum_{i<j}^{N} 
V(r_{ij}) + \sum_{i=1}^{N} U(r_i) \ ,
\label{hamiltonian}
\end{equation}
with $V(r)$ the intermolecular interaction and $U(r)$ the confining potential 
in the quasi-1D simulations. Upon the condition of moderate pressures, it is 
justified to use a radial interaction between \textit{p}-H$_2$ molecules 
because in the \textit{para} state the H$_2$ molecule is in the $J=0$ 
rotational state. We use the semi-empirical Silvera-Goldman 
potential~\cite{silvera} which has been extensively used in the past. When the 
system is not strictly 1D, we include an external potential $U(r)$ which 
confines in the radial direction. In particular, for the quasi-1D calculations 
we have worked on two cases. In a first one, we study the inner channel inside 
a (10,10) carbon nanotube of radius $R=6.80\ \text{\AA}$ coated with an 
incommensurate lattice of \textit{p}-H$_2$ of density $\sigma=0.112\ 
\text{\AA}^{-2}$. This configuration is very close to the one 
obtained in Ref.~\cite{ancilotto} for the same nanotube. In our case, we obtain 
the 
potential $U(r)$ as a sum of the interaction that an H$_2$ molecule located at 
an $r$ distance to the center would feel due to the C atoms of the nanotube and 
the H$_2$ molecules of the inert layer. At difference with other approaches 
which used the potential inside the nanotube by direct integration of the 
Lennard-Jones potential~\cite{Breton,Stan}, we include here explicitly the real 
positions of the atoms and then summed up all to give the total interaction.
The 
C-H$_2$ potential is of Lennard-Jones type, with the same parameters than in 
Ref.~\cite{gordillo2}. The second model to study the effects of departing from 
a strictly 1D geometry is a harmonic potential $U(r)=\hbar^2/(2 m 
r_0^4)\,(x^2+y^2)$, 
with $r_0$ a parameter which controls the strength of the confinement. A 
similar harmonic model was used recently in a PIMC simulation~\cite{bonintubo}.

We used $N=30$ in the major part of our simulations; partial runs with larger 
number of particles were also performed but the results were not significantly 
different, almost for the quantities of our interest. The time step was $\Delta 
\tau= 10^{-3} \ \text{K}^{-1}$ and convergence was achieved at imaginary times 
$\tau \simeq 0.25 \ \text{K}^{-1}$.

 \section{Results}
\label{sec:results}     

\begin{figure}
\begin{center}
\includegraphics[width=0.85\linewidth,angle=0]{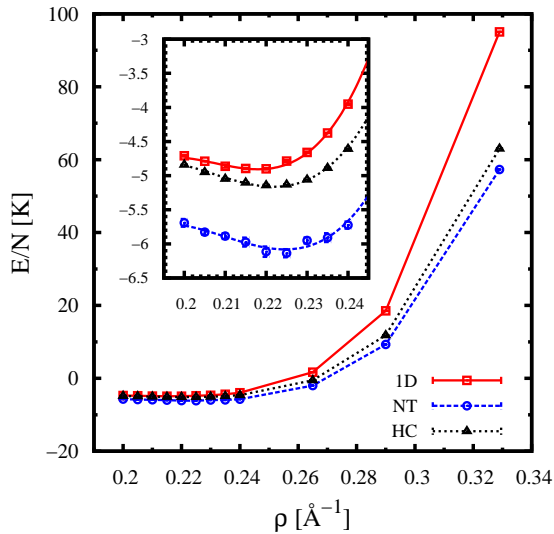}
\caption{(Color online). Energy per particle of \textit{p}-H$_2$ as a function 
of the density. The insert 
shows the same values around the equilibrium density. Harmonic case with $r_0 = 
0.51 \ \text{\AA}$.}
\label{fig:energy}
\end{center}
\end{figure}

\begin{figure}
\begin{center}
\includegraphics[width=0.85\linewidth,angle=0]{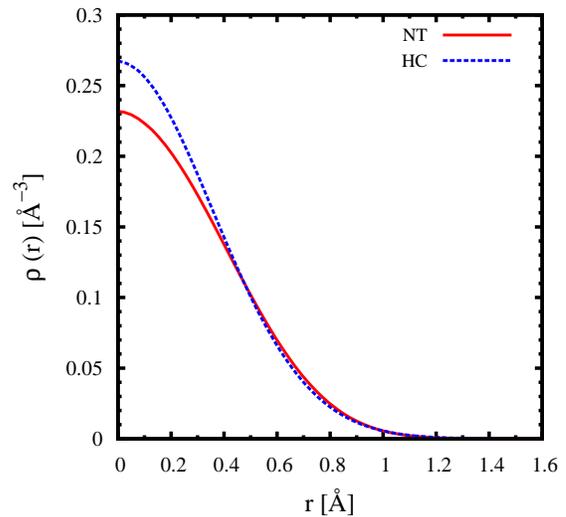}
\caption{(Color online). Radial density profile for NT and HC quasi-1D systems 
at the equilibrium density. Harmonic case with $r_0 = 0.51 \ \text{\AA}$.}
\label{fig:profile}
\end{center}
\end{figure}

The energy per particle as a function of the linear density $\rho$ is shown in 
Fig. \ref{fig:energy} for the three studied systems: 1D, (10,10) carbon 
nanotube (NT), and harmonic confinement (HC). We adjusted the parameter $r_0$ 
in the HC case to be close to the particle density profile of the NT case. By 
taking $r_0=0.51\ \text{\AA}$ we obtain in fact very similar density profiles, 
as shown in Fig. \ref{fig:profile}. Coming back to the energy results, one can 
see that near the equilibrium point the equations of state are rather similar 
(in the NT and HC cases we have subtracted to the energy per particle the 
energy of a single molecule in the same environment). The equilibrium densities 
$\rho_0$ for 1D, NT, and HC are $0.218(2)$, $0.224(2)$, and $0.221(2)\ 
\text{\AA}^{-1}$, respectively (numbers within parenthesis are the 
statistical errors).
The possibility of 
movement in the radial direction makes that the quasi-1D configuration 
equilibrates at a slightly larger density with respect to the 1D limit. On the 
other hand, the spinodal point $\rho_s$, defined as the point where the speed 
of sound becomes zero, appears in quasi-1D at densities statistically 
indistinguishable of the 1D limit, $\rho_s=0.208(3)\ \text{\AA}^{-1}$.
 However, the most significant effect of opening the 
radial direction is produced at large densities in which the growth of the 
energy with the density is clearly steeper in 1D than in the NT and HC cases.   

In 1D systems with gapless excitation spectrum, $\varepsilon(k)= \hbar k c$ 
when $k \to 0$, one can make use of the Luttinger liquid theory. This 
phenomenological theory predicts the large-distance (small momenta) of the 
distribution functions. Within this model, the results are universal in terms 
of the Luttinger parameter $K$. In a homogeneous system, like the one we are 
studying here, $K$ is determined by the Fermi velocity $v_{\text F}=\hbar 
k_{\text{F}}/m$ and the speed of sound through the relation $K=v_F/c$. In 1D, 
the Fermi momentum is $k_{\text{F}}= \pi \rho$. The estimation of $K$ for 
different densities requires of a full many-body calculation since the speed of 
sound depends strongly on the intermolecular interaction.

According to Luttinger theory~\cite{Haldane,Cazalilla,Giamarchi}, the pair 
distribution function in one dimension behaves at large distances as
\begin{equation}
g(z)=1-\frac{K}{2(k_{\text F} z)^2} + \sum_{l=1}^\infty A_l 
\frac{\cos(2 l k_{\text F} z)}{|k_{\text F} z |^{2 l^2 K}} \ ,
\label{pairdist}
\end{equation}
that is a sum of oscillating terms modulated by a power-law decaying amplitude. 
The exponents of the attenuation are only dependent on the Luttinger parameter 
$K$, whereas the amplitudes $A_l$ of each term of the sum are not determined 
within 
the Luttinger theory. The oscillations in $g(z)$ (\ref{pairdist}) can produce 
divergences at momentum values $k=2l k_{\text F}$. This can be observed in the 
static structure factor $S(k)=\langle \hat{\rho}(k) \hat{\rho}(-k)\rangle$, 
with $\hat{\rho}(k)=\sum_i \exp(-i k z)$. In fact, the height of the $l$ peak 
in $S(k)$ is given by
\begin{equation}
S(k=2l k_{\text F})= A_l N^{1-2 l^2 K} \ ,
\label{peaksk}
\end{equation}
which diverges with the number of particles $N$ for values $K<1/(2l^2)$. In 
particular, the first peak diverges when $K<1/2$. In Luttinger theory this 
regime is termed quasicrystal for the resemblance to Bragg peaks in two and 
three dimensions. However, a \textit{true} crystal in 2D and 3D shows real 
Bragg peaks in which the height of the peak increases linearly with $N$ whereas 
in 1D this only happens when asymptotically $K \to 0$.

\begin{figure}
\begin{center}
\includegraphics[width=0.85\linewidth,angle=0]{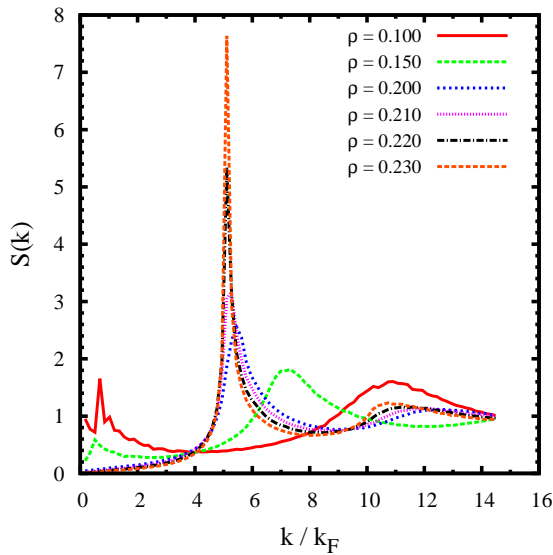}
\caption{(Color online). Static structure factor for one-dimensional H$_2$ 
at different densities (in \AA$^{-1}$).}
\label{fig:sk_1D}
\end{center}
\end{figure}

\begin{figure}
\begin{center}
\includegraphics[width=0.85\linewidth,angle=0]{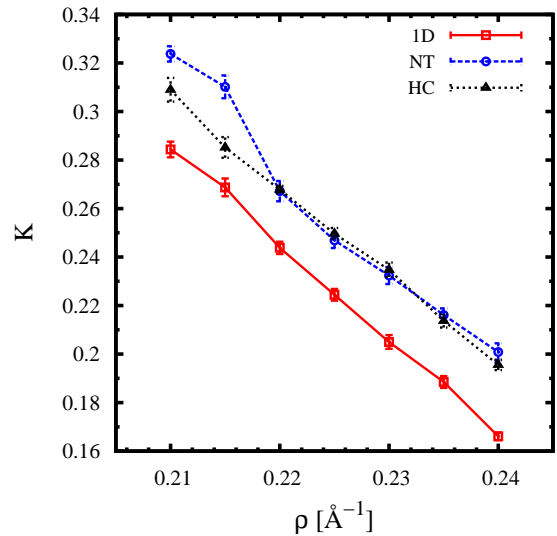}
\caption{(Color online). Luttinger parameter $K$ for the three systems under 
study as a function of the density. }
\label{fig:lutt}
\end{center}
\end{figure}

In Fig. \ref{fig:sk_1D}, we report results for the static structure factor 
$S(k)$ at different densities. From the low-$k$ linear behavior of $S(k)$ we 
can obtain the speed of sound $c$,
\begin{equation}
S(k \to 0) = \frac{\hbar k}{2mc} \ ,
\label{sk0}
\end{equation}
and, from it, the Luttinger parameter $K$. The dependence of $K$ with the 
density is shown in Fig. \ref{fig:lutt}. It has a value $K\simeq 0.25$ at the 
equilibrium density and decreases monotonically with $\rho$. The spinodal point 
is quite close to $\rho_0$ and thus 1D  \textit{p}-H$_2$ remains always in the 
quasicrystal regime. The limit of stability of the homogeneous phase, signaled 
by the spinodal point, is clearly shown in the results of $S(k)$ contained in 
Fig. \ref{fig:sk_1D}. As one can see, below the spinodal, and when $k \to 0$, 
the static structure factor shows an anomalous behavior, the linear behavior 
is lost, and the signal of a divergence is observed. Snapshots of 
configurations generated along the PIGS runs also show this break of 
homogeneity.

\begin{figure}
\begin{center}
\includegraphics[width=0.85\linewidth,angle=0]{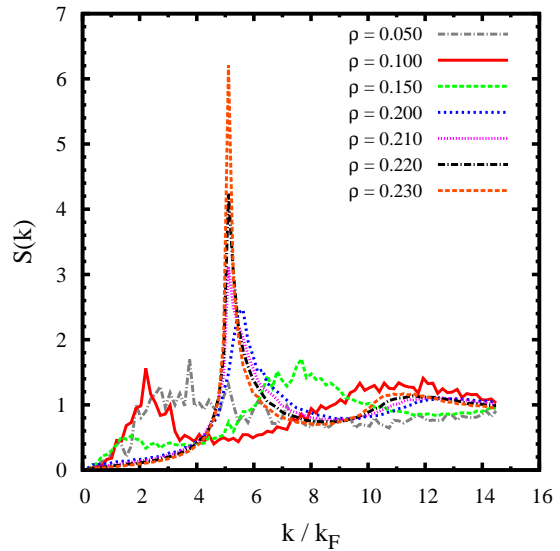}
\caption{(Color online). Static structure factor for quasi-1D H$_2$ in 
the NT case at different densities (in \AA$^{-1}$). }
\label{fig:sk_tube}
\end{center}
\end{figure}

\begin{figure}
\begin{center}
\includegraphics[width=0.85\linewidth,angle=0]{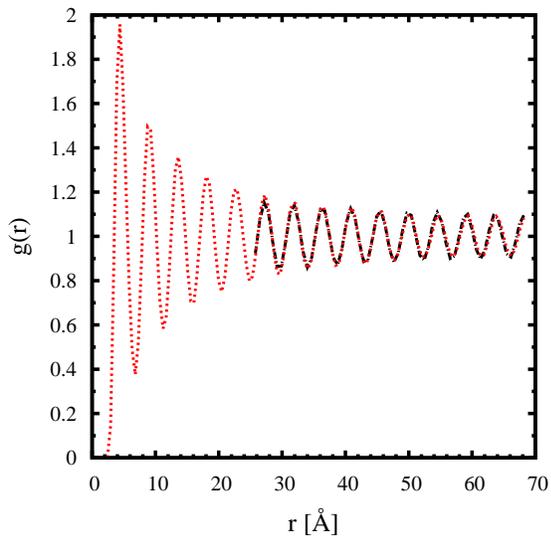}
\caption{(Color online). Two-body distribution function for the NT 
configuration at the equilibrium density. The red line is the PIGS result and 
the black one corresponds to 
the asymptotic behavior predicted by Luttinger theory (\ref{pairdist}) 
with $K$ derived from the low$-k$ behavior of $S(k)$.}
\label{fig:pairdist}
\end{center}
\end{figure}

Results for $S(k)$ along the $z$-direction in the quasi-1D NT case are shown in 
Fig. \ref{fig:sk_tube}. Above the spinodal point, the behavior of $S(k)$ is 
very similar to the purely 1D case shown in Fig. \ref{fig:sk_1D}, with a clear 
linear phononic behavior when $k \to 0$. From this behavior we estimate the 
speed of sound and the Luttinger parameter $K$. One can check that the 
Luttinger liquid theory applies to this quasi-1D system by checking if the 
asymptotic behavior of the computed $g(z)$ is well reproduced by Eq. 
(\ref{pairdist}) using the $K$ value obtained from the low $k$ linear behavior 
of $S(k)$. As one can see in Fig. \ref{fig:pairdist}, the agreement with 
Luttinger theory is excellent, confirming our premises. 

Results obtained for $K$ as a 
function of the density are shown in Fig. \ref{fig:lutt}. At equal linear 
density, the $K$ values in the NT configuration are  systematically larger than 
in purely 1D due to the effective reduction of the intermolecular interaction 
produced by the opening of radial movements. However, it still remains $K<1/2$, 
i.e., in the quasicrystal regime. When the density is lowered below the 
spinodal point the system breaks its homogeneity. As in the previous analyzed 
1D case, this instability is clearly shown in the results of $S(k)$  
(Fig. \ref{fig:sk_tube}). In spite of having larger statistical noise than in 
1D, due to the radial degree of freedom, one can see as the linear $k$ 
dependence at low $k$ is lost and a tendency to divergence is observed.

\begin{figure}
\begin{center}
\includegraphics[width=0.85\linewidth,angle=0]{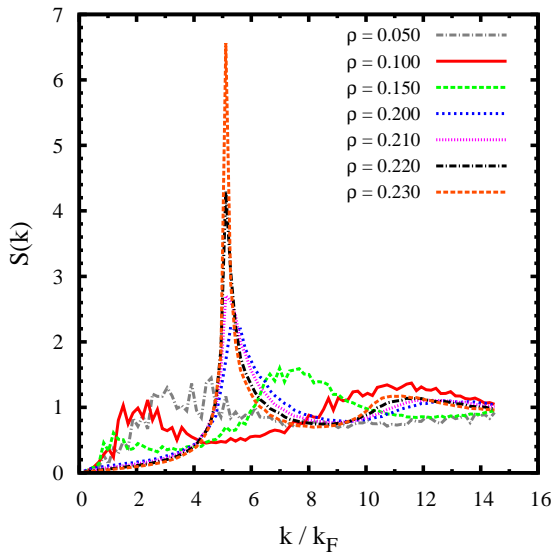}
\caption{(Color online). Static structure factor for quasi-1D H$_2$ with 
harmonic confinement HC with  $r_0 = 0.51 \ \text{\AA}$ at different 
densities (in \AA$^{-1}$).}
\label{fig:sk_hc}
\end{center}
\end{figure}

A similar analysis has been carried out in the case of a quasi-1D system with 
harmonic confinement HC. The PIGS results for $S(k)$ are shown in Fig. 
\ref{fig:sk_hc} at several densities. The observed behavior is quite close to 
the NT case since the density profiles in both cases are very similar (Fig. 
\ref{fig:profile}). One observes the break of homogeneity at the spinodal point 
and the results for $K$ in this case are also very similar to the NT case. 
These are shown in Fig. \ref{fig:lutt}; close to the equilibrium density $K$ in 
HC is slightly smaller than in NT but then both results converge to common 
values when the density grows.

\begin{figure}
\begin{center}
\includegraphics[width=0.85\linewidth,angle=0]{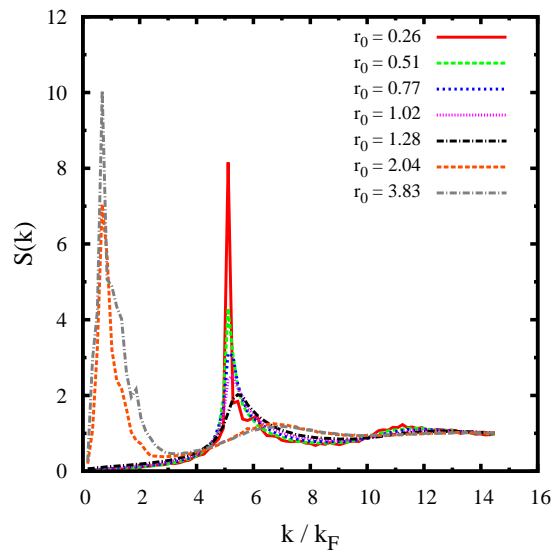}
\caption{(Color online). Static structure factor for the quasi-1D H$_2$ 
system with harmonic confinements (HC) of different strength $r_0$ (in \AA), at 
a  fixed density $\rho = 0.22$ \AA$^{-1}$.}
\label{fig:sk_ho_eq}
\end{center}
\end{figure}

The quasi-1D results for the Luttinger parameter show an enhancement of its 
value with respect the purely 1D geometry. An interesting question is to know 
if this increase could be even larger if one releases slightly the radial 
confinement, producing setups that depart more from the 1D constraint. We have 
explored this possibility by tunning the strength $r_0$ of the harmonic 
confinement HC. In Fig. \ref{fig:sk_ho_eq}, we show results of $S(k)$ for the 
HC model at a fixed density $\rho=0.22\ \text{\AA}^{-1}$ and varying the 
parameter $r_0$ in the range $0.26$-$3.83$ \AA. When the Gaussian potential is 
narrow enough, $r_0 \le 1.28$ \AA, the static structure factor is very similar 
to the 1D case, with a linear slope at low $k$ and with a strength of the peak 
decreasing slightly with $r_0$. However, when $r_0 \ge 2$ \AA $\ S(k)$ shows an 
anomalous behavior, with a main peak located at very small $k$. This reflects 
that the system breaks its homogeneity. In fact, we observe in snapshots of 
the simulations as the system aggregates in clusters of larger density. 

\begin{figure}
\begin{center}
\includegraphics[width=0.85\linewidth,angle=0]{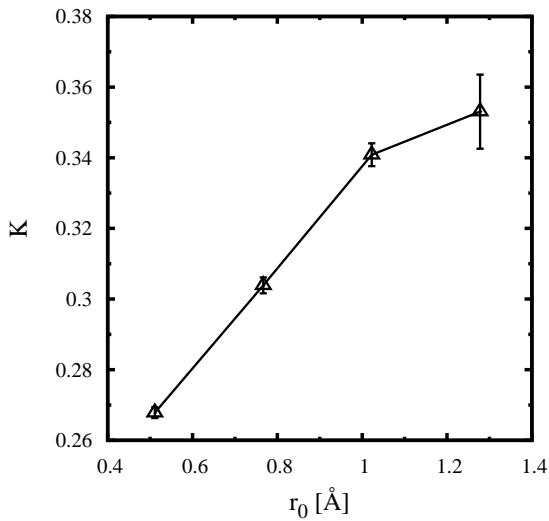}
\caption{(Color online). 
Luttinger parameter of the quasi-1D HC model at density $\rho = 0.22$ 
\AA$^{-1}$ as a function of the strength of the confinement $r_0$. 
}
\label{fig:lutt_ho}
\end{center}
\end{figure}

In Fig. \ref{fig:lutt_ho}, we show results for $K$ within the HC model as a 
function of the strength of the confinement $r_0$. They are obtained at the 
same linear density $\rho=0.22\ $ \AA$^{-1}$ and within the $r_0$ range 
in which the system is stable. We observe a linear increase of $K$ with $r_0$ 
up to $r_0 \simeq 1\ \text{\AA}$ and then it tends to flatten. At 
$r_0=1.28\ 
\text{\AA}$, we obtain 
$K=0.35$ a value which is significantly larger than in 1D at the same 
density ($K=0.25$), but still below the threshold for reaching the 
quasi-superfluid regime.

 \section{Conclusions}
\label{sec:conclusions} 

By means of the path integral ground state Monte Carlo method we have studied 
the ground-state (zero temperature) properties of 1D and quasi-1D 
\textit{p}-H$_2$. For the quasi-1D case we have used two models: the inner 
channel inside a (10,10) carbon nanotube coated with H$_2$ and a radial 
harmonic confinement with variable strength. The calculation of the equations 
of state in the three cases has allowed for an accurate determination of the 
equilibrium densities of the three systems. As expected, $\rho_0$ increases 
slightly when radial direction opens because the strong H$_2$-H$_2$ interaction 
is effectively reduced. The effect is however quite small. The spinodal point 
of the three problems is indistinguishable within our numerical resolution and 
remains very close to $\rho_0$.

From the low-$k$ behavior of the static structure factor we estimate the speed 
of sound, and from it, the Luttinger parameter $K$. In this way, we report 
results for the evolution of $K$ with the density. $K$ decreases monotonically 
with $\rho$ in all cases. In all the density regime in which the system is 
stable, $K<1/2$ and thus, according to Luttinger theory  \textit{p}-H$_2$ is a 
quasi-crystal. For a particular density, we observe as $K$ increases going from 
strictly 1D to quasi-1D but the effect is not large enough to surpass the 
quasi-crystal threshold.

Our results disagree with the recent findings of Ref.~\cite{ancilotto} using 
the same (10,10) nanotube studied here. In that work, the quasi-1D problem is 
mapped onto a purely 1D one by constructing an effective potential built from 
the obtained density profiles. The resulting effective potential is more than 
three times less attractive than the Silvera-Goldman H$_2$-H$_2$ potential and 
thus their calculation leads to the possibility of getting the quasi-superfluid 
regime because the spinodal density of this effective model is much smaller. 
Our present results show that this effective potential is not very realistic 
since a full quasi-1D calculation, accounting for the real interactions C-H$_2$ 
and 
H$_2$-H$_2$ shows that the system breaks close to the equilibrium density by 
spinodal decomposition~\cite{delmaestro}.

\begin{acknowledgments}
This research was supported by MINECO (Spain) Grants No. FIS2014-56257-C2-1-P 
and FIS2014-56257-C2-2-P.
\end{acknowledgments}

\end{document}